\documentclass[floats,preprint,prl,aps]{revtex4} 
\usepackage{graphicx}

\newcommand\beq{\begin{equation}}
\newcommand\eeq{\end{equation}}
\newcommand\bea{\begin{eqnarray}}
\newcommand\eea{\end{eqnarray}}
\newcommand{\nonum}{\nonumber}
\begin{document}

\title{\bf Fractionally Quantized Cooper Pair Stair Case in Superconducting    
Quantum Dots}

\author{\bf Sujit Sarkar}
\address{\it PoornaPrajna Institute of Scientific Research,
4 Sadashivanagar, Bangalore 5600 80, India.\\    
}
\date{\today}

\begin{abstract}
We study clean superconducting quantum dots (SQD)
and also site dependent Josephson couplings, on site charging energies and
the intersite interactions in presence of gate voltage. 
We predict the existence of different fractionally
quantized Cooper pair stair case with many interesting physical properties. The 
appearance of stair case is not only due to the Coulomb blocked
phenomena but also for the site dependent Josephson couplings. We also
explain physically the absence of other fractionally quantized Cooper pair 
stair case. The physics of fractionally quantized Cooper pair stair case 
has close resemblance with the fractionally quantized magnetization
plateau of a spin chain system under a magnetic field.

Keyword's: Superconducting quantum dots, Fractionally quantized plateau,
Abelian bosonization and Renormalization group study

\end{abstract}

\maketitle


\section{ 1. Introduction}
In last few decades mesoscopic physics of nanoscale superconducting systems
are revealing many interesting properties. There has been intense research
aimed at developing the superconducting flux based digital electronics and
computers \cite{lik1,lik2}. 
The single Cooper pair box (SCB) and single Cooper pair transistors
have developed experimentally and used to demonstrate the quantization of
Cooper pair on a small superconducting island \cite{lik3}, 
which is the foundation of
charge qubit \cite{schon1}.\\

Here we mention briefly the mechanism of
Cooper pair quantization for SCB (=
Josephson junction circuit consisting of a
small superconducting island connected via a Josephson tunnel junction to
a large superconducting island). 
The SCB was first experimentally realized by Lafarge $et~al.$ \cite{lafa}
observing the
Coulomb stair case with the step of $2 e$. The evidence of Coulomb stair
case has also predicted in Ref.\cite{bou1}. Realization of the first charge qubit
by manipulation of SCB and the observation of Rabi oscillation was done by
Nakamura $et~al$ \cite{naka1}. 
Our operating point is in the
charge regime, i.e. $ {E_c} >> {E_J}$. We consider a finite numbers of 
Cooper pairs (n) in the mesoscopic island. The eigen state equation of SCB
is
\beq
{E_c} {( \hat{n} - {n_g})}^2 |n > = {E_n} |n>
\eeq 
$n= 0,1,2$ corresponds to the charge state with energy spectrum 
${E_n} = {E_c}{( \hat{n} - {n_g})}^2 $. $n_g $ is the gate voltage
induced charge in SQD.
From Eq. 1, it is clear to us for a specific value of gate voltage 
(i.e ${n_g} =1/2 $) the charge state
$|0>$ and $|1>$ become degenerate. Switching on a small Josephson coupling
will then lift the degeneracy and forming a two level system. So the system
reduced to the qubit system.
\beq
H_{SCB} = -1/2 ( B_z {\sigma}_z  + B_x {\sigma}_x  ),
\eeq
where $ {B_z} = {E_c} (1 - 2 n_g )$ and $ {B_x} = {E_J}$. The qubit level
energies are given by the equation
$ E_{1,2}
 = ~\pm (1/2) \sqrt{ {{E_c}^2}  {(1- 2 {n_g})}^2 + {E_J}^{2} } $.
At the charge degeneracy point, ${n_g} = 1/2$, at this point only the off
diagonal part contribute and the energy levels are separated by $E_J$ and the
qubit eigen states
$ |E_1 > = |0 > - | 1>$ and $ |E_2 > = |0 > + | 1>$. For these states, the
average charge on the island is zero, while it changes to $\pm 2e$ from the 
degeneracy point, 
where the qubit eigen state approach the pure charge state.\\

It is clear to us from the analyses of previous
studies that the Cooper pair stair case 
(charge quantization state of SCB) only occurs due to presence of
Coulomb blocked phenomena. Here we consider a model systems consists of
array of clean superconducting quantum dots (SQD)  
and also a SQD systems with site dependent Josephson couplings, on-site
charging energies and the intersite interactions. We will study different
fractionally quantized Cooper pair stair case. It is clear to us from
Eq. 1 that the basic physics of SCB can be understood by the spin chain model
Hamiltonian under a magnetic field. We will see from the analysis of our
model that the factionally quantized Cooper pair stair case is nothing 
but the fractionally quantized
magnetization plateau state of spin chain system.     

\section{2. Cooper Pair Stair Case and Physical Analysis of Cooper 
of Clean
Superconducting Quantum Dots} 
The model Hamiltonian of SQD system has different parts,
\beq
H~=~H_{J1}~+~H_{J2}~+~H_{EC0}~+~H_{EC1}~+~H_{EC2}.
\eeq
We recast the different parts of the Hamiltonian in Quantum 
Phase model.\\
$
H_{J1}~=~ -E_{J1} \sum_{i} cos ({\phi}_{i+1} -{\phi}_{i})
$,
$
H_{J2}~=~ -E_{J2} \sum_{i} cos ({\phi}_{i+2} -{\phi}_{i})
$.\\
Hamiltonians $H_{J1}$ and $H_{J2}$ are Josephson energy Hamiltonians
respectively for nearest neighbor (NN) and 
next-nearest-neighbor (NNN) Josephson tunneling between the SQD. 
$
H_{EC0}~=~ \frac{E_{C0}}{2} \sum_{i}  
{(-i \frac{\partial}{{\partial}{{\phi}_i}} - \frac{N}{2})^{2} } 
$,
$
H_{EC1}~=~ E_{Z1} \sum_{i} {n_i}~{n_{i+1}} , 
$
$ H_{EC2}~=~ E_{Z2} \sum_{i} {n_i}~{n_{i+2}} .$  
${H_{EC0}}, {H_{EC1}}$, and $H_{EC2}$ are respectively the
Hamiltonians for on-site, NN and NNN charging energies of the SQD.
In the phase representation,  
$(-i \frac{\partial}{{\partial}{{\phi}_i}})$ is the operator
representing the number of Cooper pairs at the ith dot, 
and thus it takes only the integer
values ($n_i$). Hamiltonian $H_{EC0}$ accounts for the
influence of gate voltage ($e N \sim V_g$). 
$e N$ is the average dot charge induced by the gate voltage.
When the
ratio $\frac{E_{J1}}{E_{C0}} \rightarrow 0$, the SQD 
array is in the insulating state having a gap of the width
$\sim {E_{C0}}$, since it costs an energy $\sim E_{C0}$
to change the number of pairs at any dot. The exception are the 
discrete points at $N~=~2n+1$, where a dot with charge $2ne$
and $2 (n+1) e$ has the same energy because gate charge 
compensates the charges of extra Cooper pair in the dot.
On this degeneracy point, a small amount of Josephson coupling 
leads the system to the superconducting state. We are interested
in analyzing the phases explicitly near this degeneracy point.\\
Now we want to recast our basic Hamiltonians in the spin 
language. During this process we follow Ref. (\cite{lar}).
$
H_{J1}~=~ -2~E_{J1} \sum_{i}  
( {S_i}^{\dagger} {S_{i+1}}^{-} + h.c)
$,
$
H_{J2}~=~ -2~E_{J2} \sum_{i} ( {S_i}^{\dagger} {S_{i+2}}^{-} + h.c)
$
$
H_{EC0}~=~ \frac{E_{C0}}{2} \sum_{i} 
{(2 {S_i}^{Z} - h )^{2} }.$ 
$$
H_{EC1}~=~4 E_{Z1} \sum_{i} {S_i}^{Z}~{S_{i+1}}^{Z},  
$$
$$
H_{EC2}~=~4 E_{Z2} \sum_{i} {S_i}^{Z}~{S_{i+2}}^{Z} . 
$$
Here $h= \frac{(N - 2n -1)}{2}$ allows tuning of the
system to a degeneracy point by means of gate voltage.
The phase diagram is periodic in $N$ with period 2. 
Here we only consider a single slab of the phase diagram,
$0 \leq N \leq2$.
One can express
spin chain systems to spinless fermions systems through 
the application of Jordan-Wigner transformation.
In Jordan-Wigner transformation
the relations
between the spin and the electron creation and annihilation operators are
\cite{gia1}
\bea
S_i^z =  \psi_i^{\dagger} \psi_i - 1/2 ~,
S_i^-  =  {(-1)}^{i} \psi_i ~\exp [i \pi \sum_{j=-\infty}^{i-1} n_j]~,
S_i^+  =  {({S_i^-})}^{\dagger},
\label{jor}
\eea
where $n_j = \psi_j^{\dagger} \psi_j$ is the fermion number at site $j$.
Therefore,
\begin{center} 
\beq
{H}_{J1} = - 2 {E_{J1}} ~\sum_i  
~(\psi_{i+1}^{\dagger} \psi_i + 
\psi_i^{\dagger} \psi_{i+1}),
\eeq 
\beq
H_{J2} = -2 {E_{J2}} ~\sum_i ~( \psi_{i+2}^{\dagger} \psi_i + {\rm h.c.})
(\psi_{i+1}^{\dagger} \psi_{i+1} - 1/2),
\eeq
\end{center}
\begin{center}
\beq 
 H_{EC0}~=~ - 2 h E_{C0} ~\sum_i  
~(\psi_i^{\dagger} \psi_i - 1/2),
\eeq
\beq 
 H_{EC1}~=~ 4 E_{Z1} ~\sum_i (\psi_i^{\dagger} \psi_i - 1/2)
(\psi_{i+1}^{\dagger} \psi_{i+1} - 1/2) ,
\eeq
\beq 
H_{EC2}~=~ 4 E_{Z2} ~\sum_i ~(\psi_i^{\dagger} \psi_i - 1/2)
(\psi_{i+2}^{\dagger} \psi_{i+2} - 1/2) .
\eeq
\end{center}
In order to study the continuum field theory of these Hamiltonians, 
we recast the spinless
fermions operators in terms of field operators by this relation \cite{gia1}. 
\beq
{\psi}(x)~=~~[e^{i k_F x} ~ {\psi}_{R}(x)~+~e^{-i k_F x} ~ {\psi}_{L}(x)]
\eeq
where ${\psi}_{R} (x)$ and ${\psi}_{L}(x) $ describe the second-quantized 
fields of right- and 
the left-moving fermions respectively. 
$k_F$ is Fermi wave vector. 
It reveals from Eq. 7 that the applied external gate voltage on the
dot systems appears as a magnetic field in spin chain. So the different
values of applied gate voltage promote the system in the different
state of magnetization. 
In our system $k_F$ will depend on the
applied gate voltage.
The
Fermi wave vector and the magnetization state ($m$) are related
by the relation
$ {k_F} = \frac{\pi}{2} (1 - 2 m) $. 
We would like to study the effect of gate voltage in our study
so we keep $k_F$ as arbitrary. 
In our study, we will mainly
focus on the magnetization state with $m = 0, 1/4, 1/2 $. 
We would like to express the fermionic fields in terms of bosonic 
field by the relation 
\beq
{{\psi}_{r}} (x)~=~~\frac{U_r}{\sqrt{2 \pi \alpha}}~~e^{-i ~(r \phi (x)~-~ \theta (x))},
\eeq
$r$ is denoting the chirality of the fermionic fields,
right (1) or left movers (-1).
The operators $U_r$ commutes with the bosonic field. $U_r$ of different species
commute and $U_r$ of the same species anti-commute. 
$\phi$ field corresponds to the 
quantum fluctuations (bosonic) of spin and $\theta$ is the dual field of $\phi$. 
They are
related by the relations $ {\phi}_{R}~=~~ \theta ~-~ \phi$ and  $ {\phi}_{L}~=~~ \theta ~+~ \phi$.
The model Hamiltonian after continuum field theoretical studies for arbitrary
values of $k_F$ is
\bea
H & = & {H_0} + ~\frac{4 E_{Z1}}{{(2 \pi {\alpha})}^2}
 \int : cos( 4 \sqrt{K} {\phi } (x)~- (G - 4 k_F)x -2 k_F a ):~ dx \nonum\\ 
& & + ~\frac{4 E_{Z2}}{{(2 \pi {\alpha})}^2}
 \int : cos( 4 \sqrt{K} {\phi } (x)~+ (G - 4 k_F)x - 4 k_F a ):~ dx 
\eea
\bea
H_0 & = &  ( \frac{v}{2 \pi} + \frac{8 E_{C0}}{{\pi}^2} - 4 E_{J2}
- \frac{2 {E_{J1}}^2}{E_{C0}}) 
 ~\int dx ~[ 
:{{({{\partial}_x} \theta)}^2}:
+ :{{({{\partial}_x} \phi)}^2}:~] \nonum\\
& & + (16 E_{C0} - 8 E_{J2} - 4 \frac{{E_{J1}}^2}{E_{C0}}) 
  \int dx~
: ({\partial}_x {\theta} - {\partial}_x {\phi}) 
({\partial}_x {\theta} + {\partial}_x {\phi}): 
\eea
$H_0 $ is the non-interacting part of the model Hamiltonian, 
$v$ is the velocity of low energy excitations
liquid parameter and the
other is  
$ K~ ( =~\sqrt{[\frac{ E_{J1}~-~ \frac{32}{\pi} E_{J1} E_{Z2} }
{E_{J1}~+~\frac{2}{\pi} (4 E_{Z1}~ -~\frac{3 {E_{J1}}^2 }{4 E_{C0}})}]}) $.

\section{2.1 Calculations and Results for  
Cooper Pair Stair Case for Single Pair in Alternate Sites
( $m=0$ Magnetization Plateau)
 } 
In this sub-section we discuss the Cooper pair stair case for a single
pair in alternate sites.  
We have already mentioned that the SQD array system is nothing
but the spin-1/2 chain and stair case of the SQD is nothing but the magnetization
plateau.
We will see that at $ m=0$ (${k_F} = \pi/2 $) there is an evidence of magnetization
plateau. We would like to mention  
basic criteria for the appearance of magnetization plateau before we start our 
full swing discussion on stair case physics. 
The basic criteria for the appearance of plateau is the following:
It is well known to us that the elementary excitation of an one
dimensional spin system is gapless for half integer spin chain and
gapped for integer spin chain.
In the presence of magnetic field, it is possible
for an integer spin chain to be gapless with partial magnetization and 
a half-odd-integer
spin chain to show a gap above the ground state for appropriate values of the
magnetic field. It has been shown by different groups \cite{oshi,tot,shi,tone,kole}
that the magnetization of some systems
can exhibit plateaus at certain nonzero values for some finite ranges of 
the magnetic
field.
The basic criteria for the appearance of magnetization plateau can be 
understood from
the
extension of Lieb-Schultz-Mattis theorem under a magnetic field. This 
implies that translationally
invariant spin chains in an applied field can be gapful without 
breaking translation
symmetry
under the condition $S-m=$ integer, where $S$ and $m$ are the spin and the
magnetization state of the chain. In this gapful phase magnetization 
plateau occurs for
quantized
values of m. Fractional quantization can also occur, if accompanied by 
spontaneous breaking
of
translational symmetry. Fractional quantization can be understood 
from the $S-m=$ non
integer
condition. In this situation system is either in the gapless low lying 
states or the
degenerate
ground state with spontaneous translational symmetry breaking 
in the thermodynamic limit.
These conditions for the
appearance of plateau are the necessary but not the sufficient condition. 
The nature
and the
occurrence of the plateau are also dependent on the nature of interaction present
in the system \cite{cab,mutt}.

Our model Hamiltonian for $m=0$ magnetization plateau, in terms of bosonic fields, are
\beq                                                          
H = H_0 ~-~\frac{4 ( E_{Z1} - E_{Z2} )}{{(2 \pi {\alpha})}^2}  
 \int cos( 4 \sqrt{K} {\phi } (x)~)~ dx 
\eeq  
In the spin chain, system is in the Neel-Ising state
when XXZ anisotropy
is greater than 1 ($ 4 E_{Z1}  > 1 $). 
This state is doubly degenerate state and $S-m=1/2$
, so this phase of the system satisfy the all criteria of fractionally
quantized magnetization plateau. This is the relevant
physics when the NNN exchange interaction is less than a critical
value. When NNN exchange interaction exceed some critical value,
ground state of the system is dimerized and doubly degenerate. The dimerized
ground state is the product of spin singlet of adjacent sites \cite{sar,hal}.
In both ordering the lattice translational symmetry is breaking. 
In our clean SQD
system, we have predicted the CDW wave state
with lattice translational symmetry breaking is
alike to Neel-Ising state of
one dimensional spin systems. This CDW state is the one-dimensional
counter part of the two-dimensional checker board phase.
One
can find the dimer order density wave for the lower values of Cooper
pair density, when one consider
further long range interaction between
the SQD. To get the dimer-order density in the presence of NNN interactions
between dots, we shall have to consider
the charge degeneracy point at higher values of half-integer
Cooper pair density like
$n~=~3/2$. This Cooper pair stair case is originating due to the 
intersite Coulomb repulsion, so it is related with the Coulomb
blocked phenomena of the system.  

\section{2.2 Calculations and Results for 
Cooper Pair Stair Case for Single Pair in Every Two Sites
($m=1/4$ Magnetization Plateau) }
 
In this sub-section we discuss the occurrence of finite 
magnetization plateau ($ m= 1/4$).
This plateau state correspondence to the Cooper pair stair case for single
Cooper pair in every two sites. For weak dimerization, i.e.,
when there is no reduction of unit cell size, reciprocal lattice
vector is $ 2 \pi$ then the Hamiltonian of the system reduced to
Eq. 15. None of the sine-Gordon coupling terms will provide the
significant contributions   
for this plateau state,
due to the oscillatory nature of the integrand.
Effective Hamiltonian of the system for this state is the following:
\bea
H &=& H_0 ~-~\frac{4 E_{Z1}}{{(2 \pi {\alpha})}^2}
 \int (-1)^{x}~cos( 4 \sqrt{K} {\phi } (x)~)~ dx \nonum\\
& & + ~\frac{4 E_{Z2}}{{(2 \pi {\alpha})}^2}
\int (-1)^{x}~cos( 4 \sqrt{K} {\phi } (x)~)~ dx ~ 
\eea
For strong dimerization state, reciprocal lattice vector $G$ reduced to
$\pi$ due to the reduction of the unit cell. 
The model Hamiltonian of the system reduced to

\beq                                                          
H = H_0 ~-~\frac{4  E_{Z12}}{{(2 \pi {\alpha})}^2}  
 \int cos( 4 \sqrt{K} {\phi } (x)~+~\beta)~ dx,
\eeq 
where $ E_{Z12} ~=~ \sqrt{ {E_{Z1}}^2 ~ +~{E_{Z2}}^2 }$ and 
$\beta ~= tan^{-1} \frac{E_{z2}}{E_{Z1}} $.
To study the renormalization
group flow diagram, we shall have to construct the renormalization
group (RG) equations.
The RG equations of $H_2$ is
\bea
\frac{d K(l)}{dl}~ &=~& - {E_{Z12}}^2 {K(l)}^2 \nonum \\
\frac{d E_{Z12}}{dl} ~&=~& (2 - 4 K(l)~) ~E_{Z12}
\eea
In Fig. 1, we present the renormalization group flow diagram.
RG flow lines, flowing off to
$E_{Z12} \rightarrow~\infty$
are an
Ising-Neel state, an anti-ferromagnetic ordering is
setup in the z direction, i.e., the system is in the
CDW wave state with only one boson in every two sites.
This is the, B1 phase in the RG flow diagram. When the RG
flow lines, flowing off to
$E_{Z12} \rightarrow~-\infty$ are the dimer order density
waves. In this state two Cooper pair of adjacent sites
are bound with each other. This is the insulating state of the
system. In spin chain, dimerized ground state is the analogous
state of this phase. This phase is depicted as a B2 phase in
the RG flow diagram. Phase D is the repulsive Luttinger liquid
phase, i.e., the gap-less phase
due to the irrelevancy of sine-Gordon coupling terms. Phase E is
the superconducting phase of the system.
This Cooper pair stair case is originating due to the
intersite Coulomb repulsion, so it is related with the Coulomb
blocked phenomena of the system.

\section{ 2.3 Calculations and Results for 
Cooper pair stair case at empty band limit
($m=\frac{1}{2}$ Magnetization Plateau):}

Now we discuss the saturation plateau at $m= \frac{1}{2}$ ($k_F ~=~0$).
$K_F ~=~0$ implies that the band is empty and the dispersion is not linear, 
so the validity of the continuum field theory is questionable. Values of the two Luttinger 
liquid parameters, $v_0$ and $K$, are $0$ and $1$ respectively.                 
It also implies that none of the sine-Gordon coupling terms become              
relevant in this parameter space.                                                                          
Saturation plateaus are only appearing due to very high values of magnetic field. 
In this plateau, system is in ferromagnetic ground  state and restore 
the lattice translational symmetry.                           
We think, this is the classical phase of the system.                            

Absence of other fractionally quantized magnetization plateaus: 
Here we present the possible explanation for the absence of other fractionally
quantized magnetization plateaus (like $\frac{1}{3},\frac{1}{5}$ etc):
A careful examination of Eq. 12 reveals that to get a non oscillatory
contribution from  Hamiltonian
one has to be satisfied $4 k_F=G$ condition but this condition is not fulfilled for these
plateaus. The integrand of this sine-Gordon coupling terms contain an
oscillatory factor that leads to a vanishing contribution. The other criteria is
that non vanishing sine-Gordon coupling term should be relevant.

%
%
\section{3. Model Hamiltonian and Physical Analysis of Cooper Pair Stair 
Case of Site Dependent Couplings, 
Superconducting Quantum Dots}

At first we write down the model Hamiltonian of SQD system with site
dependent Josephson couplings, on-site charging energies and intersite
interactions in presence of gate voltage.
\beq
H~=~H_{J1}~+~H_{J2}~+~H_{EC0}~+~H_{EC1}~+~H_{EC2}.
\eeq
We recast the different parts of the Hamiltonian in Quantum
Phase model.\\
\begin{center}
$
H_{J1}~=~ -E_{J1} \sum_{i} (1 - {(-1)^i} {\delta}_1) cos ({\phi}_{i+1} -{\phi}_{i})
$,
$
H_{J2}~=~ -E_{J2} \sum_{i} cos ({\phi}_{i+2} -{\phi}_{i})
$.\\
\end{center}
Here NN Josephson couplings are different
on alternate sites.
${E_{J1}} (1 + {\delta}_1 )$ and
${E_{J1}} (1 - {\delta}_1 )$ are the Josephson coupling strength
for odd and even site respectively.
\begin{center}
$
H_{EC0}~=~ \frac{E_{C0}}{2} \sum_{i} (1 - (-1)^{i} {\delta}_2 )
{(-i \frac{\partial}{{\partial}{{\phi}_i}} - \frac{N}{2})^{2} }
$,
\end{center}
$ E_{C0} ( 1+ {\delta}_2 )$ and
$ E_{C0} ( 1- {\delta}_2 )$ are the on site charging energies for
odd and even site respectively.
$$
H_{EC1}~=~ E_{Z1} \sum_{i} (1 - (-1)^{i} {\delta}_3 ) {n_i}~{n_{i+1}},
$$
$ E_{Z1} ( 1+ {\delta}_3 )$ and
$ E_{Z1} ( 1- {\delta}_3 )$ are the inter-site charging energies for
odd and even site respectively. All $\delta$'s are the deviational
parameter from the clean superconducting quantum dots.
$$ H_{EC2}~=~ E_{Z2} \sum_{i} {n_i}~{n_{i+2}} ,$$
${H_{EC0}}, {H_{EC1}}$, and $H_{EC2}$ are respectively the
Hamiltonians for on-site, NN and NNN charging energies.
Following the prescription of previous sections, one can write the continuum
model Hamiltonian.
\bea
H &=& {H_0} - 2 {E_{J1}} \int dx : sin ( 2 \sqrt{K}{\phi} (x) - (2 k_F - \pi)x ):\nonum\\ 
& &  +~\frac{4 E_{Z1}}{{(2 \pi {\alpha})}^2} 
 \int : cos( 4 \sqrt{K} {\phi } (x)~- (G - 4 k_F)x -2 k_F a ):~ dx \nonum\\
& & + \frac{4 E_{Z1} {\delta}_3 }{{(2 \pi {\alpha})}^2}
 \int (-1)^{x} : cos( 4 \sqrt{K} {\phi } (x)~- (G - 4 k_F)x -2 k_F a ):~ dx \nonum\\
& &  + \frac{2 h E_{C0}}{\pi \alpha} \int (-1)^{x} 
: sin( 2 \sqrt{K} {\phi } (x) +2 k_F x ):~ dx
 \nonum\\
& & + ~\frac{4 E_{Z2}}{{(2 \pi {\alpha})}^2}
 \int : cos( 4 \sqrt{K} {\phi } (x)~+ (G - 4 k_F)x - 4 k_F a ):~ dx
\eea

\section{3.1 Calculation and Results for Cooper Pair stair case of a site 
dependent SQD array for 
single Cooper pair in alternate sites ($m=0$ magnetization plateau)}

In this sub section, we find the evidence of Cooper pair stair case for
single pair in alternate site for site dependent SQD system.
The effective Hamiltonian reduce to 
\bea
H & = & {H_0} + 2 E_{J1} {{\delta}_1} \int ~dx :cos(2 \sqrt(K) \phi (x)): 
+ 2 \frac{h E_{C0}}{\pi \alpha} {{\delta}_2} 
\int ~dx :cos(2 \sqrt(K) \phi (x)): \nonum\\
& & - \frac{4 E_{Z12}}{{(2 \pi \alpha)}^2} \int ~dx :cos(4 \sqrt{K} \phi (x)): 
\eea
At around the charge
degeneracy point, the system is in the mixed state ($M$), i.e.,
the simultaneous presence of dimer density wave and
staggered phase.
Our model Hamiltonian consists of three sine-Gordon couplings.
The first one arises due to site dependent variations of NN Josephson
couplings, it yields the dimerized phase of the system.
The anamolous scaling dimension of this term is 2$K$.
This phase
is spontaneous, i.e., infinitesimal variation of NN Josephson coupling
is sufficient to produce this state.
We have predicted in the previous sections of this work that the stair
case is appearing due to the Coulomb blocked phenomena. Here we observe
that the stair case may also occur for the Josephson tunneling.
This prediction was absent in the previous studies.
This observation of dimerized
phase is in contrast with the clean superconducting quantum dots array,
where it appears when NNN
Josephson coupling exceed some critical value. The second sine-Gordon
coupling term arises due to the site dependent applied gate voltage.
The anamolous scaling dimension of this term is 2$K$.
It yields the staggered phase of the system. The system is in the mixed
phase ($M$) when both of the couplings are in equal magnitude otherwise the
system is in any one of the state of this mixed phase, depending on the strength of
couplings.
We have ignored the 3rd sine-Gordon coupling term of the Hamiltonian,
because the anamolous scaling dimension is much larger than the
other two.

\section{3.2 Calculation and Results for Cooper Pair stair case of a site 
dependent SQD array for 
single Cooper pair in every two sites ($m= 1/4$ magnetization plateau) and
empty band limit ($m =1/2 $ magnetization plateau) }

In this sub section, we find the evidence of Cooper pair stair case for
single pair in every two sites and also for empty band limit of site dependent 
SQD system.\\

$m =1/4$:
in the weak dimerization limit, Hamiltonian reduced to
\beq
H  =  {H_0} 
- \frac{4 E_{Z1}{~{\delta}_3 }}{{(2 \pi \alpha)}^2} 
\int ~dx :sin(4 \sqrt{K} \phi (x)): 
\eeq
The sine-Gordon coupling term became relevant when $K < 1/2$. If we analyze
the expression for Luttinger liquid parameter then we get the following 
quadratic equation of the system parameters to achieve this stair case phase
for weak dimerization limit.
$$
{E_{J1}}^2 - E_{J1} (4 - \frac{512}{3} {E_{Z1}} ) - 4 E_{Z1} \leq 0 
$$
This phase is in contrast with the stair case phase of the homogeneous SQD.
In weak dimerization, there is no evidence of stair case phase for homogeneous
SQD system. This phase of the system satisfies the all criteria of fractional
quantization.
Now we discuss
for stronger strength of dimerization, i.e., when
the dimerization strength exceed some critical value, at this point
reciprocal lattice vector
$G$ reduced from $2 \pi$ to $\pi$. The Hamiltonian of the system
reduced to 
\beq
H  = {H_0} 
-  \frac{4 E_{Z12}}{{(2 \pi \alpha)}^2} \int ~dx :cos(4 \sqrt{K} \phi (x)+ \beta):
\eeq
Expression for $E_{Z12}$ and $\beta$ has given in 
section 2.2. This equation is identical to Eq. 16. 
So the physics of the system is the same as
Fig. 1. Hence the physical behavior of the system in strong dimerization 
limit is the same for homogeneous and site dependent SQD.\\

$m=1/2$ and other fractionally quantized plateaus: In these phases non of the 
sine-Gordon coupling terms are relevant.
So there is no plateau phases for this states of the system.
Hence the physical behavior of the system in strong dimerization
limit is the same for homogeneous and site dependent SQD.\\

\begin{figure} 
\includegraphics[scale=0.50,angle=0]{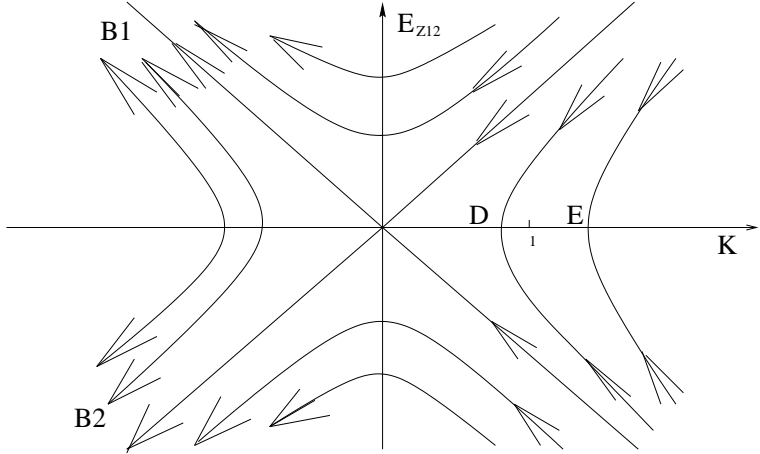} 
\caption{Renormalization group flow diagram of
the SQD array. We
have depicted the different phases of the model Hamiltonians by:
B1. Charge-density wave (CDW), 
B2. Dimer-order density wave;
D. Second kind of Repulsive Luttinger liquid; 
E. Superconducting phase. 
} 
\label{Fig. 1 }
\end{figure}
\section{4. Conclusions}
We have predicted the Cooper pair stair case phenomena for homogeneous
and also for the site dependent SQD system. We have predicted 
interesting physics for different stair case. We also conclude that
the appearance of Cooper pair stair case is not only related with the
Coulomb blocked phenomena but also related with the Josephson junction
tunneling.

The author would like to acknowledge 
the Center for Condensed Matter Theory
of IISc for providing the working space.

\end{document}